\documentclass{moriond}

\usepackage{units}
\usepackage{url}
\usepackage{amsfonts}

\def\be{\begin{equation}}
\def\ee{\end{equation}}
\def\bea{\begin{eqnarray}}
\def\eea{\end{eqnarray}}

\newcommand{\Photo}{\includegraphics[height=35mm]{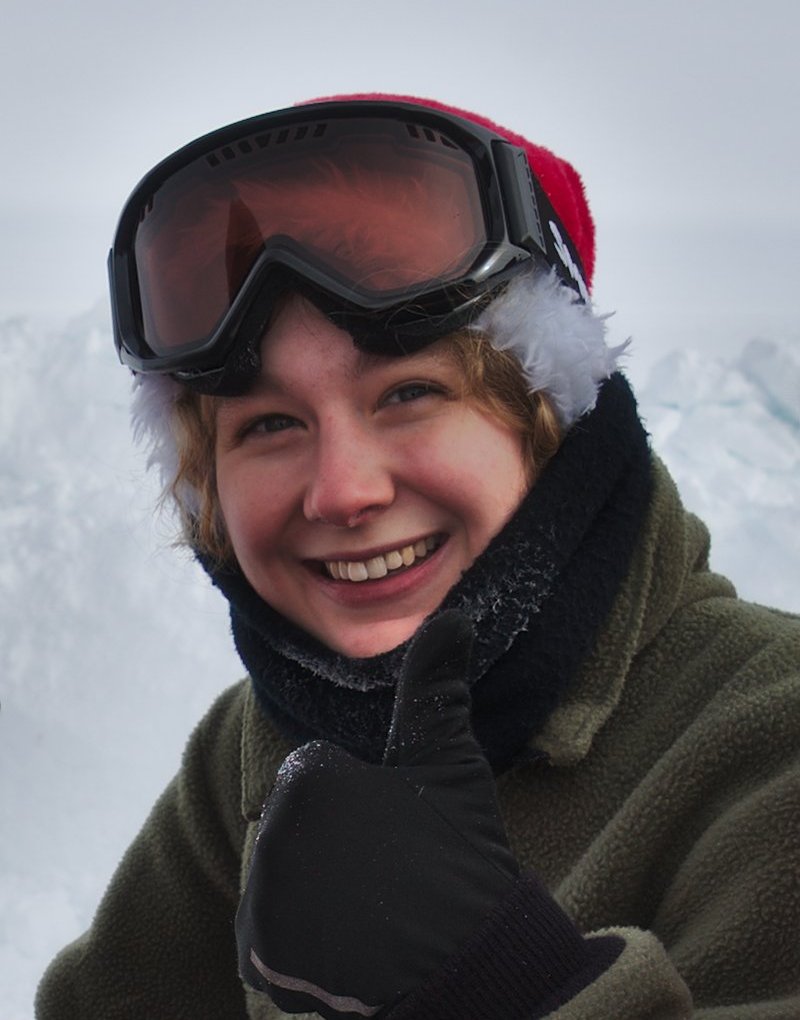}}

\begin{document}
\vspace*{4cm}
\title{NEUTRINO CONSTRAINTS TO SCOTOGENIC DARK MATTER INTERACTING IN THE SUN}

\author{T. de Boer$^{\dagger}$, R. Busse$^{\ddagger,*}$, A. Kappes$^{\ddagger}$, M. Klasen$^{\dagger}$, S. Zeinstra$^{\dagger}$}

\address{
$^{\dagger}$Institut für Theoretische Physik/ $^{\ddagger}$Institut für Kernphysik, Westfälische Wilhelms-Universität Münster,\\Wilhelm-Klemm-Str. 9, 48149 Münster, Germany
}

\maketitle\abstracts{
Radiative seesaw models have the attractive property of providing dark matter candidates in addition to generation of neutrino masses. Here we present a study of neutrino signals from the annihilation of dark matter particles which have been gravitationally captured in the Sun, in the framework of the scotogenic model. We compute expected event rates in the \textsc{IceCube} detector in its 86-string configuration. As fermionic dark matter does not accumulate in the Sun, we study the case of scalar dark matter, with a scan over the parameter space. Due to a naturally small mass splitting between the two neutral scalar components, inelastic scattering processes with nucleons can occur. We find that for small mass splittings, the model yields very high event rates. If a detailed analysis at \textsc{IceCube} can exclude these parameter points, our findings can be translated into a lower limit on one of the scalar couplings in the model. For larger mass splittings only the elastic case needs to be considered. We find that in this scenario the XENON1T limits exclude all points with sufficiently large event rates. 
}

\section{Introduction}

While there is strong evidence for the existence of large amounts of dark matter in the universe, the nature of dark matter is still unknown. Radiative seesaw models, like the famous scotogenic model~\cite{Ma06}, present an interesting solution to this problem, while at the same time only extending the Standard Model (SM) by very few fields. 

In the framework of the scotogenic model, WIMPs (weakly interacting massive particles) can accumulate in large celestial bodies like the Sun. The WIMPs can loose energy by scattering off nuclei inside the Sun, until they are gravitationally captured and sink into the core, producing a local over-density which results in a boost of self-annihilation. The WIMPS can either annihilate directly into neutrinos or into other SM particles, which in turn decay into neutrinos. This produces a flux that can be measured by neutrino telescopes such as \textsc{IceCube}~\cite{Aart16}.

The accumulation of dark matter in the Sun is directly linked to the WIMP-nucleon scattering cross section. In this work, we consider both elastic and inelastic scattering. The latter arises through a natural mass splitting of the two neutral components in the case of scalar dark matter in the scotogenic model. We investigate the parameter space by means of a numerical scan, in order to identify models that yield sufficiently large neutrino event rates in \textsc{IceCube} in its \textsc{ic86} string configuration, and could therefore be studied in future \textsc{IceCube} analyses.

\section{The scotogenic model}

The scotogenic model extends the Standard Model by two new fields: One fermion singlet $N_{i}$ with three generations, and one scalar doublet with the components $\left(\eta^{+},\eta^{0}\right)$. An additional $\mathbb{Z}_{2}$ symmetry, under which the new fields are odd and all Standard Model fields are even, ensures the stability of the lightest odd particle by preventing its further decay. The lightest odd particle is the dark matter candidate of the scotogenic model, provided it is neutral~\cite{Ma06}. 

The new terms to the Lagrangian are:
\begin{eqnarray}
\mathcal{L}_{N} & = & -\frac{m_{N_{i}}}{2}N_{i}N_{i}+y_{i\alpha}\left(\eta^{\dagger}L_{\alpha}\right)N_{i}+\mathrm{h.c.}-V,
\end{eqnarray}
\begin{eqnarray}
V & = & m_{\phi}^{2}\phi^{\dagger}\phi+m_{\eta}^{2}\eta^{\dagger}\eta+\frac{\lambda_{1}}{2}\left(\phi^{\dagger}\phi\right)^{2}+\frac{\lambda_{2}}{2}\left(\eta^{\dagger}\eta\right)^{2}+\lambda_{3}\left(\phi^{\dagger}\phi\right)\left(\eta^{\dagger}\eta\right)\nonumber \\
 &  & +\lambda_{4}\left(\phi^{\dagger}\eta\right)\left(\eta^{\dagger}\phi\right)+\frac{\lambda_{5}}{2}\left[\left(\phi^{\dagger}\eta\right)^{2}+\left(\eta^{\dagger}\phi\right)^{2}\right]
\end{eqnarray}
where $m_{N_{i}}$ is the mass matrix of the fermion singlet, the
$L_{\alpha}$ are the left-handed Standard Model lepton doublets with
$\alpha=1,2,3$ generations, $y_{i\alpha}$ is a Yukawa coupling
matrix where $i=1,2,3$ are the generations of the new right-handed
neutrinos, the $\lambda_{1-5}$ are the new coupling parameters, and $\phi$ is the Standard Model Higgs field. 

After electroweak symmetry breaking, the squared masses of the new physical scalar bosons are given by 
\begin{eqnarray}
m_{\eta^{+}}^{2} & = & m_{\eta}^{2}+\lambda_{3}\langle \phi^{0}\rangle^{2},\nonumber \\
m_{\eta^{0R}}^{2} & = & m_{\eta}^{2}+\left(\lambda_{3}+\lambda_{4}+\lambda_{5}\right)\langle \phi^{0}\rangle^{2},\label{eq: scalar masses}\\
m_{\eta^{0I}}^{2} & = & m_{\eta}^{2}+\left(\lambda_{3}+\lambda_{4}-\lambda_{5}\right)\langle \phi^{0}\rangle^{2},\nonumber 
\end{eqnarray}
where $\langle\phi^{0}\rangle=\unit[246.22/\sqrt{2}]{GeV}$ is the Higgs vacuum expectation value~\cite{Zyla:2020zbs}. The real and imaginary components of the neutral scalar $\eta^{0}$, indicated by the superscripts $R$ and $I$, respectively, obtain slightly different masses. This mass splitting $\delta$ is governed by the coupling parameter $\lambda_{5}$, which must be naturally small, since if $\lambda_5$ is exactly zero, the neutrinos would be massless and lepton number would be conserved, leading to a larger symmetry. For bounds on the other model parameters, please refer to Ref.~\cite{deBoer:2021pon}.

\section{WIMP-nucleon scattering in the Sun}

\begin{figure}
\begin{minipage}{1.0\textwidth}
  \centering
  \raisebox{-0.5\height}{\includegraphics[width=0.35\textwidth]{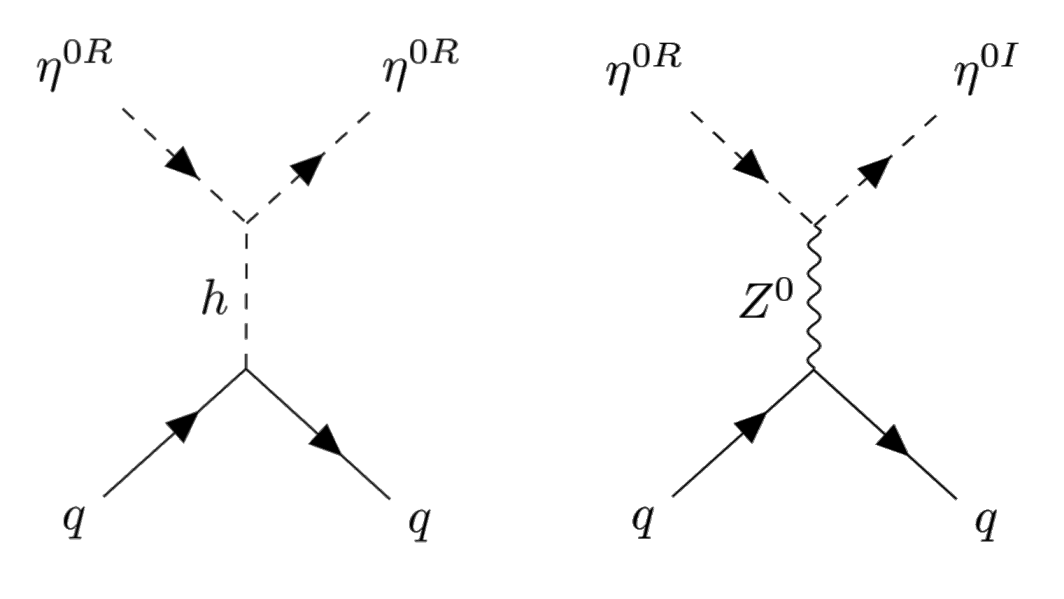}}
  \hfill
  \raisebox{-0.5\height}{\includegraphics[width=0.6\textwidth]{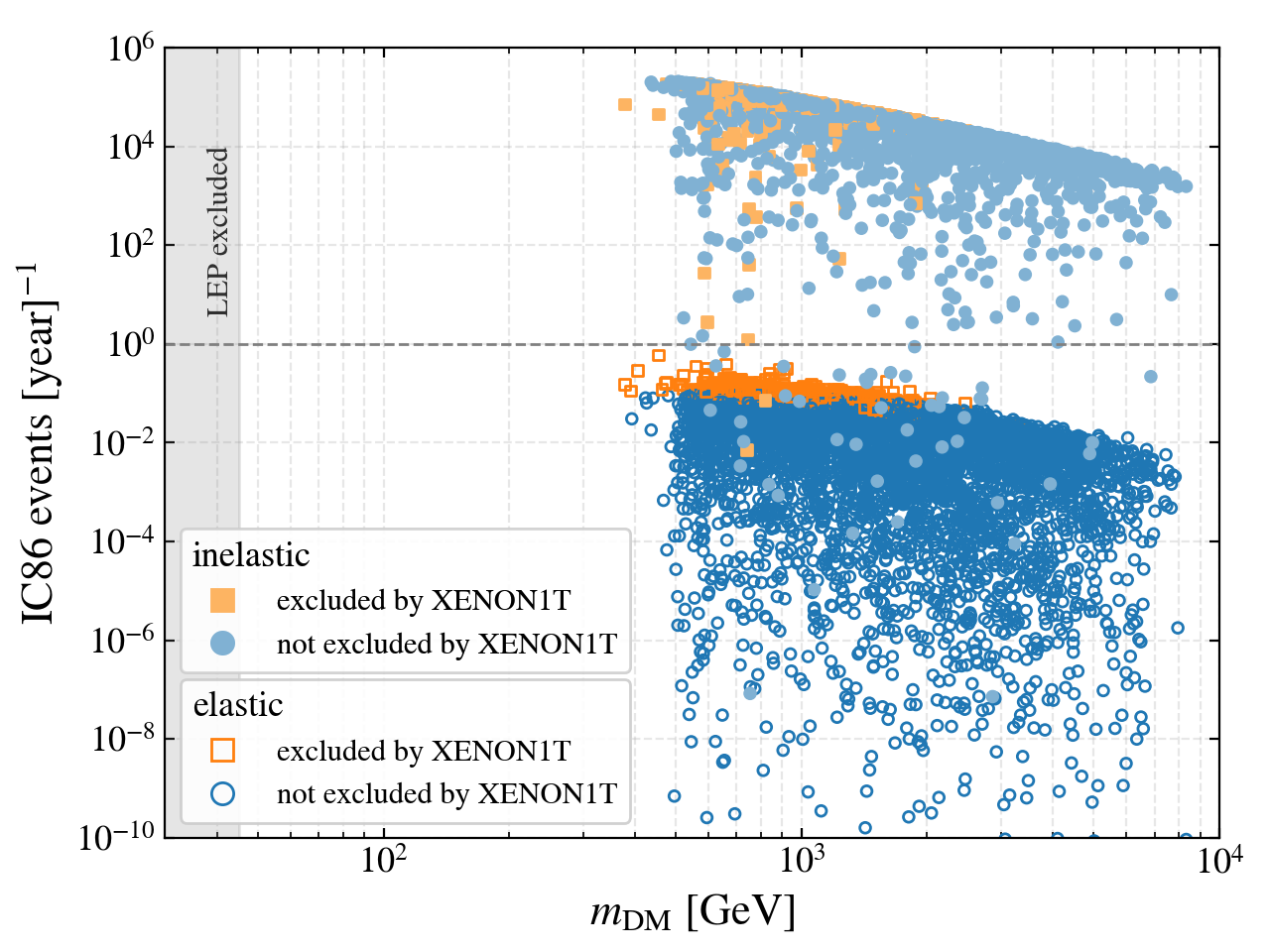}}
\end{minipage}
\caption{\label{tollesbild}
    \textbf{Left:} Feynman diagrams of the elastic and inelastic scalar dark matter-nucleon scattering processes in the scotogenic model. In this case the dark matter candidate is $\eta^{0R}$. For the case where $\eta^{0I}$ is the dark matter candidate, one can replace $\eta^{0R}$ with $\eta^{0I}$ and vice versa. 
    \textbf{Right:} Event rate in \textsc{ic86} vs. the mass of the dark matter candidate. Models with a small mass splitting between $\eta^{0R}$ and $\eta^{0I}$, for which inelastic scattering dominates, are marked with filled points. Models with larger mass splittings are marked with hollow points. For both cases, orange points are excluded by XENON1T, while blue points are not. The grey dashed line marks one \textsc{ic86} event per year. The grey shaded area marks masses that are excluded by the LEP limit from the invisible $Z^0$ width.
    }
\end{figure}

In contrast to fermion dark matter, scalar dark matter has a non-zero
spin-independent WIMP-nucleon cross section which enables accumulation in the Sun. The diagrams for scalar dark matter scattering off of nucleons are shown in Fig. \ref{tollesbild} (left hand side). Next to elastic scattering, the existence of a slightly heavier state allows for inelastic scattering as well, where the dark matter particle transitions into the heavier state, provided that the mass splitting $\delta$ between the two states fulfills $\delta < \frac{v^2}{2}\mu$, where $\mu$ is the WIMP nucleon reduced mass and $v$ is the relative velocity~\cite{TuckerSmith:2001hy}. For small $\lambda_5$ the mass splitting can be approximated by:
\begin{equation}
    \delta \approx \frac{\lambda_5 \langle\phi^{0}\rangle^2}{ m_{\eta^{0R/I}}},
    \label{eq: del lam5}
\end{equation}
depending on the smaller of the masses $m_{\eta^{0R}}$ and $m_{\eta^{0I}}$. This shows the dependence of the mass splitting on $\lambda_5$ explicitly.~\cite{deBoer:2021pon}

The inelastic scattering off of quarks in the scotogenic model is mediated by a $Z^0$-exchange. The amplitude can be decomposed into a vector and an axial-vector part. For scalar dark matter, the axial-vector part vanishes in the non relativistic limit~\cite{Agrawal:2010fh}. The remaining vector interaction contributes to the spin-independent scattering cross section.

\noindent 
\vspace{-15pt}

\section{Results of the numerical scan}

To investigate the scotogenic model numerically, we implement the model using \textsc{Sarah 4.14.0}~\cite{Staub:2013tta}. The mass spectrum and branching ratios are then computed with \textsc{SPheno 4.0.3}~\cite{Poro03,PS11}, and observables such as the relic density and neutrino fluxes are calculated using \textsc{micrOMEGAs 5.0.8}~\cite{BBGPZ18}. Since the inelastic scenario was not implemented in \textsc{micrOMEGAs}, we use \textsc{CalcHEP 3.7}~\cite{Belyaev:2012qa} to calculate the the WIMP-quark matrix elements and the cross section in the inelastic case. The inelastic capture rate is calculated with \textsc{DarkSUSY-6.2.3}~\cite{Bringmann:2018lay} and inserted into modified \textsc{micrOMEGAs} routines. For the elastic scenario we use pre-existing \textsc{micrOMEGAs} routines for both the scattering cross section as well as the capture rate. The event rate in \textsc{ic86} is calculated with a modified routine in \textsc{micrOMEGAs}, with the latest data for the \textsc{IceCube} effective area~\cite{Aartsen:2016zhm}.

Figure \ref{tollesbild} (right hand side) shows results of our numerical scan. The \textsc{ic86} event rate is plotted against the mass of the dark matter particle. We see two groups of models: The lower group consists mainly of models with a large mass splitting between $\eta^{0R}$ and $\eta^{0I}$, for which inelastic scattering is kinematically forbidden. The upper group with significantly higher event rates consists of models with small mass splittings and dominant inelastic dark matter. Note that we show only those points that survive the following constraints: PLANCK relic density~\cite{Agha18}, the mass of the Higgs boson~\cite{Aad15}, LFV limits~\cite{Baldi16,Bell88,Dohm93}, and limits on the Standard Model neutrino masses (we use the Casas-Ibarra parametrization~\cite{CI01} to calculate the Yukawa coupling matrix). We plot also the LEP exclusion from the invisible $Z^0$ width~\cite{Cao:2007rm,Lundstrom:2008ai}, to show that it does not constrain our parameter space. The remaining models can further be constrained by direct detection limits, where the XENON1T limit on the spin-independent WIMP-nucleon cross section~\cite{Aprile:2018dbl} imposes the strongest and only constraint to our parameter space. In both groups of models we draw the XENON-excluded ones in orange, whereas in the inelastic group they are excluded for their elastic scattering cross section, since the XENON limit does not apply to inelastic dark matter. 

We see that no model in the elastic group yields \textsc{IceCube} event rates above one event per year, whereas almost all models in the inelastic group do, up to event rates of $\mathcal{O}(10^5)$ per year. The non-excluded inelastic models could therefore be probed by \textsc{IceCube}. The findings of a respective analysis can be translated into a lower limit of the coupling parameter $\lambda_{5}$.

\section{Summary}

In this work we investigated neutrino signals from WIMP annihilations in the Sun within the framework of the scotogenic model. We performed a numerical scan of the parameter space, in order to identify models that yield an \textsc{ic86} event rate above one event per year, and are viable under a number of experimental constraints. Both elastic and inelastic dark matter have been considered. We find that elastic dark matter yields \textsc{ic86} event rates of only $\mathcal{O}(10^{-1})$, while for inelastic dark matter there exist viable models with event rates up to $\mathcal{O}(10^5)$. The findings of an \textsc{IceCube} analysis probing these models could be translated into a lower limit on the coupling parameter $\lambda_{5}$ of the scotogenic model.

\section*{Acknowledgments}

This work has been supported by the BMBF and the DFG through the Research Training Group 2149 "Strong and weak interactions – from hadrons to dark matter".

\section*{References}

{\small{}  
\bibliography{mp_rbusse.bbl}
}{\small\par}

\end{document}